	\documentclass[final,number,12pt]{elsarticle}
	 \usepackage{graphicx}
	\usepackage{amssymb}
	 \usepackage{amsthm}
	 \usepackage{amsmath}
	 \usepackage{subfigure}%
	 \usepackage{float}
	 \usepackage{multirow}
	 \usepackage{multicol}
	 \usepackage{epstopdf}
	 \usepackage[english]{babel}
	 \usepackage[autostyle]{csquotes}
	  \usepackage{ifpdf}
	     \ifpdf
	       \usepackage{tikz}
	     \fi

	 \biboptions{numbers,square, sort, comma}

	\journal{--}
	
\begin{document}
		\begin{frontmatter}

	\title{Anomalous volatility scaling in high frequency financial data}

	\author[rvt,els]{Noemi Nava\corref{cor1}}
	\ead{n.morales.11@ucl.ac.uk}
	\author[focal]{T. Di Matteo}
	\ead{tiziana.di\_matteo@kcl.ac.uk }
	\author[rvt,els]{Tomaso Aste}
	\ead{t.aste@ucl.ac.uk}

	\cortext[cor1]{Corresponding author}
	
	\address[rvt]{ Department of Computer Science, University College London, Gower Street, London, WC1E 6BT, UK}
	\address[focal]{Department of Mathematics, King's College London, The Strand, London, WC2R 2LS, UK}
	\address[els]{Systemic Risk Centre, London School of Economics and Political Sciences, London, WC2A 2AE, UK}
	
\begin{abstract}
Volatility of intra-day stock market indices computed at various time horizons exhibits a scaling behaviour that differs from what would be expected from  fractional Brownian motion (fBm).
We investigate this anomalous scaling by using   empirical mode decomposition (EMD), a method which separates  time series into a set of  cyclical components at different time-scales.
By applying the EMD to fBm, we retrieve a  scaling law that relates the variance of the components  to a power law of the  oscillating period. In contrast, when analysing  22  different stock market indices, we observe  deviations from the fBm and Brownian motion scaling behaviour. We discuss and quantify these deviations, associating them  to the  characteristics of financial markets, with  larger   deviations corresponding to   less developed markets.
	 	 
\end{abstract}

	
	\end{frontmatter}

\section{Introduction}
	
Over the last few years financial markets have witnessed the availability and widespread use of data sampled at high frequencies. The study of these data allows to identify the intra-day structure of financial markets \citep{Dacorogna, Bartolozzi}. Data  at these frequencies have dynamic properties which are not generated by a single process but by several components that are superimposed on top of each other. These components are not immediately apparent, but once identified, they  can be meaningfully categorized as noise, cycles at different time-scales and trends \citep{Dacorogna}.

Since the early work of Mandelbrot \cite{Mandelbrot, Mandelbrot67},  it was recognized that different time-scales contribute to  the complexity of financial time series  in a self-similar (fractal) manner. Empirical properties of financial data at various frequencies have been observed in a number of studies, see for example \cite{Muller2,Gencay2001,Cont01,Lillo2004,Glattfelder}.

According to the  random walk hypothesis  \citep{fama},  financial market dynamics can be described by a random walk, a self-similar process with scaling  exponent (Hurst exponent) $H=0.5$ \citep{Hurst1951}. 
Opposing this theory, Peters \cite{Peters1994} introduced the  fractal market hypothesis (FMH), which represents financial market dynamics by  fractional Brownian motion (fBm), a self-similar process with scaling exponent $0<H<1$. The  focus of the FMH is  on the interaction of  agents with various investment horizons and differing interpretations of information. Based on this theory,  heterogeneous markets models have  explained some stylised facts (such as volatility clustering, kurtosis, fat tails of returns, power law behaviours) observed in financial markets, see for example \cite{LeBaron,Hommes, Chiarella, Lux}.

In self-similar uni-scaling process, such as fBm, all time-scales contribute proportionally and there is a specific  relation that links statistical properties at different time-scales \citep{Feller1971}. However, real financial time series have more complex  scaling patterns,  with some time-scales contributing disproportionally; these patterns characterize  multi-scaling processes whose  statistical properties vary at each time-scale \citep{DiMatteo,Matteo,Anirban,Barunik2012,Buonocore}.

The knowledge of scaling laws in financial data  helps us to understand market dynamics \citep{Mantegna1995, Mantegna2000}, that can be interpreted to construct efficient and profitable trading strategies. In this paper, we use empirical mode decomposition (EMD),  an algorithm introduced by  Huang \cite{Huang}, to decompose intra-day financial time series into  a trend and  a finite set of simple oscillations. 
These oscillations, called intrinsic mode functions (IMFs),  are associated with  the  time-scale of cycles latent in the time series. 
The EMD provides a tool for an exploratory analysis that takes into account both the fine and coarse structure of the data. This decomposition has been widely used in many fields, including the analysis of financial time series \citep{Huang2003,Wu2007, Guhathakurta2008, Qian, Nava2014}, river flow fluctuations \citep{Huang2009103}, wind speed \citep{Guo2012}, heart rate variability \citep{Balocchi2004}, etc.

In this paper, we first apply  EMD  to fBm, uncovering a power law scaling  between the  period and  variance of the IMFs with  scaling exponent  related to the Hurst exponent.
We then apply EMD to 22 different stock market indices whose prices  are sampled at 30~second intervals over a time span of 6 months. In this case, we encounter more complex scaling laws than in fBm.  
The deviations from the fBm behaviour are quantified and interpreted as an anomalous multi-scaling behaviour.

This paper is organized as follows. In section 2, we introduce the EMD. In section 3, we present the variance scaling properties of  fBm. In section 4, we present an application to high frequency  financial data. We finally conclude in section 5.

	\section{Empirical mode decomposition}
	\label{EMD}
	
	The empirical mode decomposition  is a fully data-driven decomposition that can be applied to non-stationary and non-linear data  \citep{Huang}. Differently from the Fourier and the wavelet transform, the EMD does not require any a priori filter function \citep{Peng2005}. The purpose of the method is to identify  a finite set of  oscillations with scale defined by the local maxima and minima of the data itself.  Each oscillation is  empirically derived from data and is referred as an intrinsic mode function.
	An IMF must satisfies two criteria:
	\begin{enumerate}
	\item The number of extrema and the number of zero crossings must either be equal or differ at most by one.
	\item At any point, the mean value of the envelope defined by the local maxima and the envelope defined by the local minima is zero. 
	\end{enumerate}
	
	The IMFs are obtained through a  process that makes use of local extrema to separate oscillations starting with the highest frequency. Hence, given a time series $x(t)$, $t=1,2,...,T$, the process decomposes it into a finite number of intrinsic mode functions   denoted as $IMF_k(t)$, $k= 1,..., n$ and a residue $r_n(t)$. 	
	If the decomposed data consist of uniform scales in the frequency space, the EMD acts as a dyadic filter and the total number of IMFs is close to n$=\log_2(T)$ \citep{Flandrin}.	
	The residue is  the non-oscillating drift of the data. At the end of the decomposition  process, the original time series can be reconstructed as:
	
	\begin{equation}
	x(t)=\sum_{k=1}^n IMF_k(t) + r_n(t).
	\label{eq:EMD}
	\end{equation}

	The  EMD comprises the following steps:
	\begin{enumerate}
	\item Initialize the residue to the original time series $r_0(t)=x(t)$ and set the IMF index $k=1$.

	\item Extract the $k^{th}$ IMF:
	
	\begin{enumerate}
	\item initialize $h_0(t)= r_{k-1}(t)$ and the iteration counter $i=1$;
	
	\item find the local maxima and local minima  of $h_{i-1}(t)$; 
	
	\item create the upper envelope $E_u(t)$ by interpolating between the maxima (lower envelope $E_l(t)$ for minima, respectively); 
	
	\item calculate the mean of both envelopes as $m_{i-1}(t) = \frac{ E_u(t) + E_l(t)}{2}$;

	\item subtract the envelope mean from the input time series, obtaining  $h_{i}(t)= h_{i-1}(t)- m_{i-1}(t)$; 
	
	\item verify if $h_{i}(t)$  satisfies the IMF's conditions:
	
	\begin{itemize}
	\item if $h_{i}(t)$ does not satisfy the IMF's conditions, increase $i=i+1$ and repeat the sifting process from step (b); 
	
	\item if $h_{i}(t)$ satisfies the IMF's conditions, set  $IMF_k(t)= h_{i}$ and define  $r_k(t)= r_{k-1}(t)-IMF_k(t)$.
	
	\end{itemize}
	\end{enumerate}
	
	\item When the residue  $r_k(t)$ is either a constant, a monotonic slope or contains only one extrema  stop the  process, otherwise continue the decomposition from step $2$ setting $k=k+1$.
	
	\end{enumerate}

Orthogonality cannot be   theoretically guaranteed, but in most cases it is  satisfied \citep{Huang}. Including the residue as the last component and rewriting Equation \ref{eq:EMD} as $x(t)= \sum\limits_{k=1}^{n+1} C_k(t)$,  the square of the values of  $x(t)$ can be expressed as: 
\begin{equation}
x(t)^2= \sum_{k=1}^{n+1} C^2_k(t) +  \sum\limits_{\substack{j=1 \\ j \neq k}}^{n+1} \sum\limits_{k=1}^{n+1} C_k(t) C_j(t).
\end{equation}

If the decomposition is orthogonal, the cross terms should be zero. An index of orthogonality (IO) is defined as \citep{Huang}:

\begin{equation}
IO = \sum_{t=1}^T \frac{\sum\limits_{\substack{j=1 \\ j \neq k}}^{n+1} \sum\limits_{k=1}^{n+1} C_k(t) C_j(t)}{x(t)^2}.
\label{eq:IO}
\end{equation}

\section{Self-similar scaling exponent}
Self-similarity or scale invariance is an attribute of many laws of nature and is the underlying concept of fractals. Self-similarity is related to the occurrence of similar patterns at different time-scales. In this sense, probabilistic properties of self-similar processes remain invariant when the process is viewed at different time-scales \citep{Mandelbrot1982, Mandelbrot1997, Calvet2002}.
		
A stochastic process $X(t)$ is statistically self-similar, with scaling  exponent $0<H<1$, if for any real $a>0$ it follows the scaling law: 
\begin{equation}
		X(at)\overset{d}{=}a^{H}X(t) \quad \quad t \in \mathbb{R},
	\end{equation} 	
where the equality ($\overset{d}{=}$) is in  probability distribution \citep{Calvet2002}. 
		
An example of self-similar process is fractional Brownian motion (fBm), a stochastic process characterized by a positive scaling exponent $0<H<1$ \cite{Mandelbrot1968}.  When $ 0<H<\frac{1}{2} $, fBm is said to be anti-persistent with   negatively auto-correlated increments. For the case $\frac{1}{2} < H < 1$, fBm reflects a persistent behaviour and the increments are positively auto-correlated. When $H = \frac{1}{2}$,  fBm is reduce to a process with independent increments known as Brownian motion.

\subsection{EMD based scaling exponent}

Flandrin et al. \cite{Flandrin04} empirically showed that when decomposing  fractional Gaussian noise (fGn), the differentiation process of fBm \citep{Mandelbrot1968}, the  EMD  can be used to estimate the scaling exponent $H$, if   $H > \frac{1}{2}$. The authors ascertained that the variance progression across IMFs satisfies,
 $var(IMF^{fGn}_k) \propto \,\tau_k^{2(H-1)}$,  where  the function $\tau_k$ denotes the period of the $k^{th}$-IMF \footnote{The periods $\tau_k$ can be approximated as the total number of data points divided by the total number of zero crossings of each IMF.}. 

In this paper, we follow a similar approach to \cite{Flandrin04}, but instead of applying EMD to fGn, we considered its integrated process, fBm.  We  empirically showed that a similar scaling law holds for the variance of IMFs: 

  \begin{equation}
  var(IMF^{fBm}_k)\propto \, \tau_k ^{2H}.              
  \label{eq:fBmScaling1}
  \end{equation}
  
 Therefore, for fBm and fGn, the IMF variance follows a power law scaling behaviour with respect to its particular period of oscillation, and the scaling parameter  is related to the Hurst exponent.  
 The EMD   estimator of $H$ can be  determined by the slope of a linear regression fit on the logarithmic  of the variance as a function of the logarithmic of the period, 
 \begin{equation}
 \log\left(var(IMF^{fBm}_k)\right)=2H \log(\tau_k)+\log(c_0),
   \label{eq:fBmScaling}
 \end{equation}
 where $c_0$ is the intercept constant of the linear regression.
 In the following section we will provide the simulation results  supporting equation \ref{eq:fBmScaling}.

\subsection{FBm simulation analysis}
\label{sec:Simulations}
In order to verify the empirical scaling law  of Equation \ref{eq:fBmScaling1}, we generated $N=100$ fBm processes for the following  values of  the  scaling exponent   $H=0.1, 0.2, 0.3, 0.4, 0.5, 0.6, 0.7, 0.8$ and $0.9$. The simulated processes have two different lengths, $T_1=10,000$ and $T_2=100,000$.\footnote{All the fBm paths were generated using MATLAB\textsuperscript{\textregistered} wavelet toolbox.}  

\begin{figure}
 \begin{center}
 \begin{minipage}{150mm}
 \subfigure{
 \resizebox*{7cm}{!}{\includegraphics{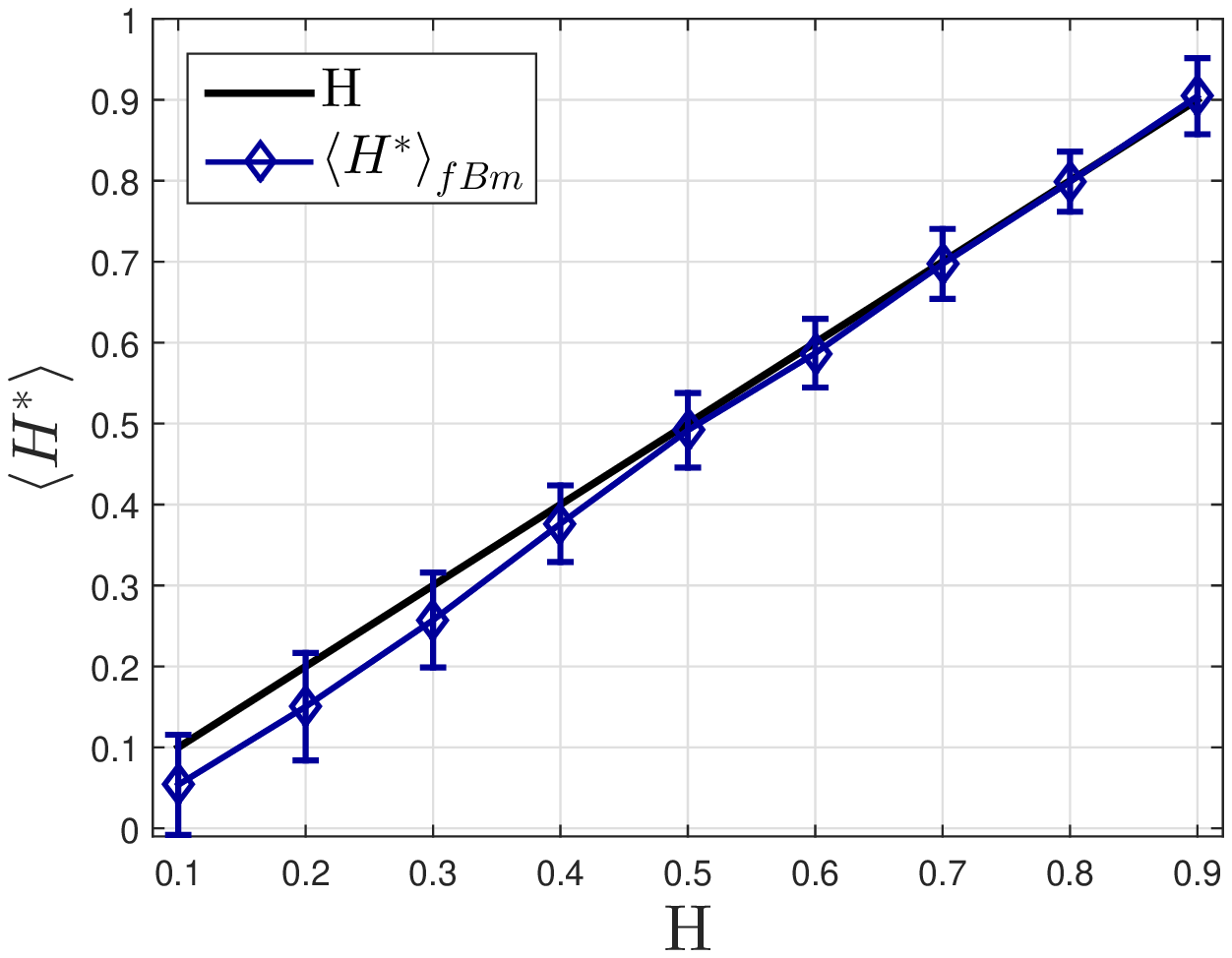}}}
 \subfigure{
 \resizebox*{7cm}{!}{\includegraphics{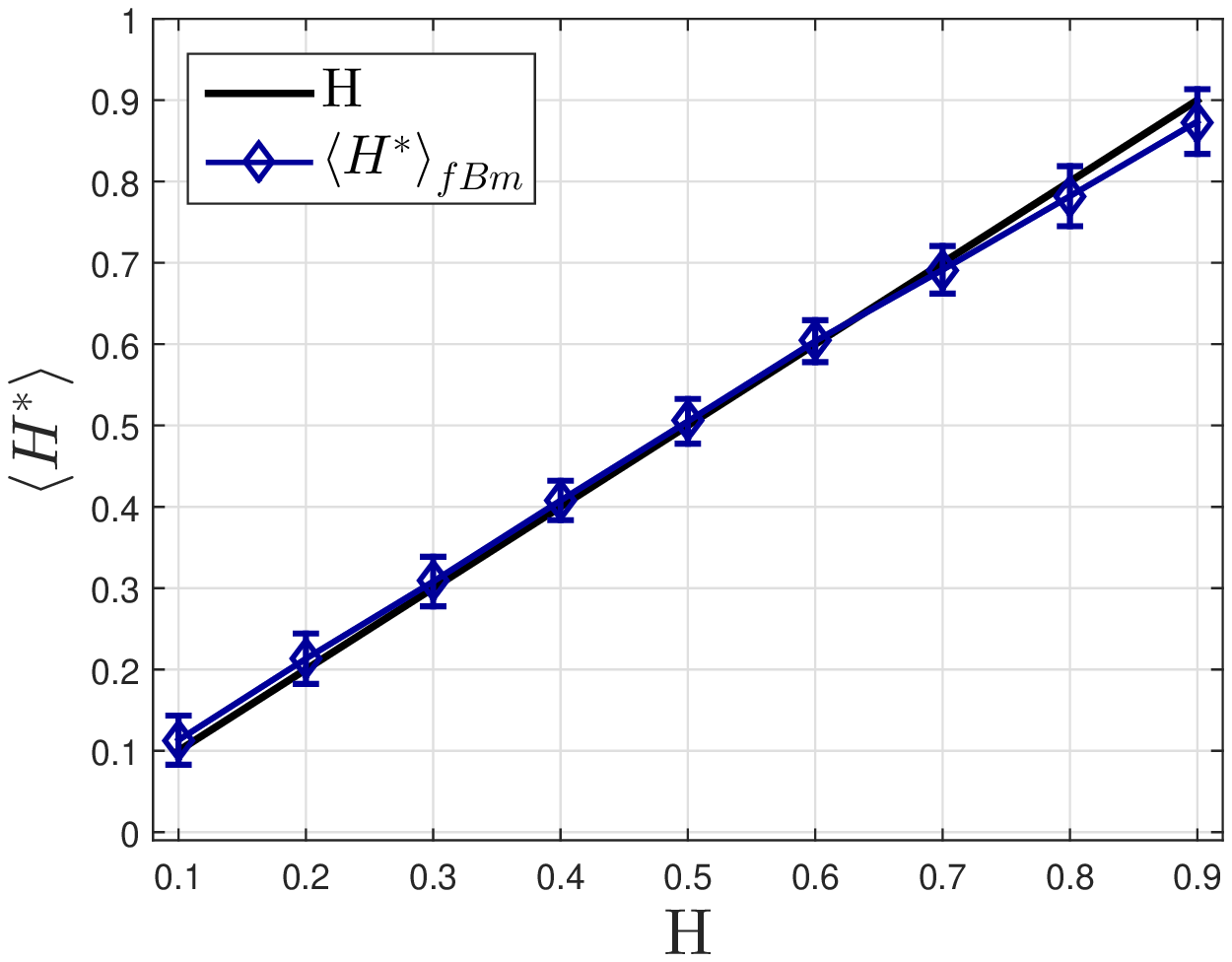}}}
   \caption{Demonstration that the empirical scaling law of Eq. 3 retrieves the expected scaling exponent for fractional Brownian motion. Mean of the scaling exponent $H^*$ over 100 simulations of fBm  with  parameter $H=0.1, 0.2, \dots, 0.9$ and length: left  $T_1=10,000$ and  right: $T_2=100,000$. The error bars denote the RMSE of the estimator.   } 
 \label{fig:HEMDB}
 \end{minipage}
 \end{center}
 \end{figure}

We applied the EMD to each fBm simulation and calculated its respective  $H^*$ exponent.  In Table \ref{tab:HEMDB}, we report $\left<H^*\right>_{fBm}$, the mean  over the 100 estimators. We also report the root mean square error (RMSE) of the estimators, $RMSE=\sqrt{\frac{\sum\limits_{i=1}^{N}\left(H_i^*-H\right)^2}{N}}$.
We observe that the longer the analysed time series, the better the estimation of $H$ is. For length $T=100,000$, $\left<H^*\right>_{fBm}$ is indeed very close to the  scaling exponent $H$ (for all values of $H$).
In Figure \ref{fig:HEMDB}, we plot the mean values of the $H^*$ exponent as presented in Table \ref{tab:HEMDB}. The error bars represent the RMSE of the estimator.
Let us emphasize that we do not propose the EMD as  a way to estimate the Hurst exponent, but as a tool to analyse the interactions between the different time scales present in the data.
For comparison, we estimated the Hurst exponent using the generalized exponent approach \cite{Matteo}. In Table \ref{tab:HEMDB},   we include the   mean and the RMSE of this estimator denoted  as $H_G$.

\begin{table}[htbp]
  \centering
  \scalebox{0.8}{
    \begin{tabular}{ccccc|cccc}
\hline
\hline
 & \multicolumn{4}{c}{\textbf{10,000}}    & \multicolumn{4}{c}{\textbf{100,000}} \\
\hline
\hline
      \boldmath{$H$}     & \boldmath{$\left<H^*\right>$} & \boldmath{$RMSE_{H^*}$} & \boldmath{$\left<H_G\right>$} & \boldmath{$RMSE_{H_G}$} & \boldmath{$\left<H^*\right>$} & \boldmath{$RMSE_{H^*}$}  & \boldmath{$\left<H_G\right>$} & \boldmath{$RMSE_{H_G}$} \\
    \textbf{0.1} & 0.05  & 0.06  & 0.15  & 0.05  & 0.11  & 0.03  & 0.15  & 0.05 \\
    \textbf{0.2} & 0.15  & 0.07  & 0.22  & 0.03  & 0.21  & 0.03  & 0.22  & 0.03 \\
    \textbf{0.3} & 0.26  & 0.06  & 0.31  & 0.01  & 0.31  & 0.03  & 0.31  & 0.01 \\
    \textbf{0.4} & 0.38  & 0.05  & 0.40  & 0.01  & 0.41  & 0.02  & 0.40  & 0.01 \\
    \textbf{0.5} & 0.49  & 0.05  & 0.50  & 0.01  & 0.51  & 0.03  & 0.50  & 0.01 \\
    \textbf{0.6} & 0.59  & 0.04  & 0.60  & 0.01  & 0.60  & 0.03  & 0.60  & 0.01 \\
    \textbf{0.7} & 0.70  & 0.04  & 0.70  & 0.01  & 0.69  & 0.03  & 0.70  & 0.01 \\
    \textbf{0.8} & 0.80  & 0.04  & 0.79  & 0.01  & 0.78  & 0.04  & 0.79  & 0.02 \\
    \textbf{0.9} & 0.90  & 0.05  & 0.88  & 0.03  & 0.87  & 0.04  & 0.87  & 0.03 \\
\hline
\hline
    \end{tabular}%
    }
    \label{tab:HEMDB}
  \caption{Confirmation that the empirical scaling law of Eq. 3 retrieves the expected scaling exponent for fractional Brownian motion. Mean of the scaling exponent $H^*$ over 100 simulations of fBm with  parameter $H=0.1, 0.2, \dots, 0.9$ and length: left  $T_1=10,000$ and  right: $T_2=100,000$. For comparison, we included the mean and the RMSE of the generalized Hurst exponent estimator denoted as $H_G$.} 
\end{table}%

\begin{figure}
    \begin{center}
  \includegraphics[width=3.5 in]{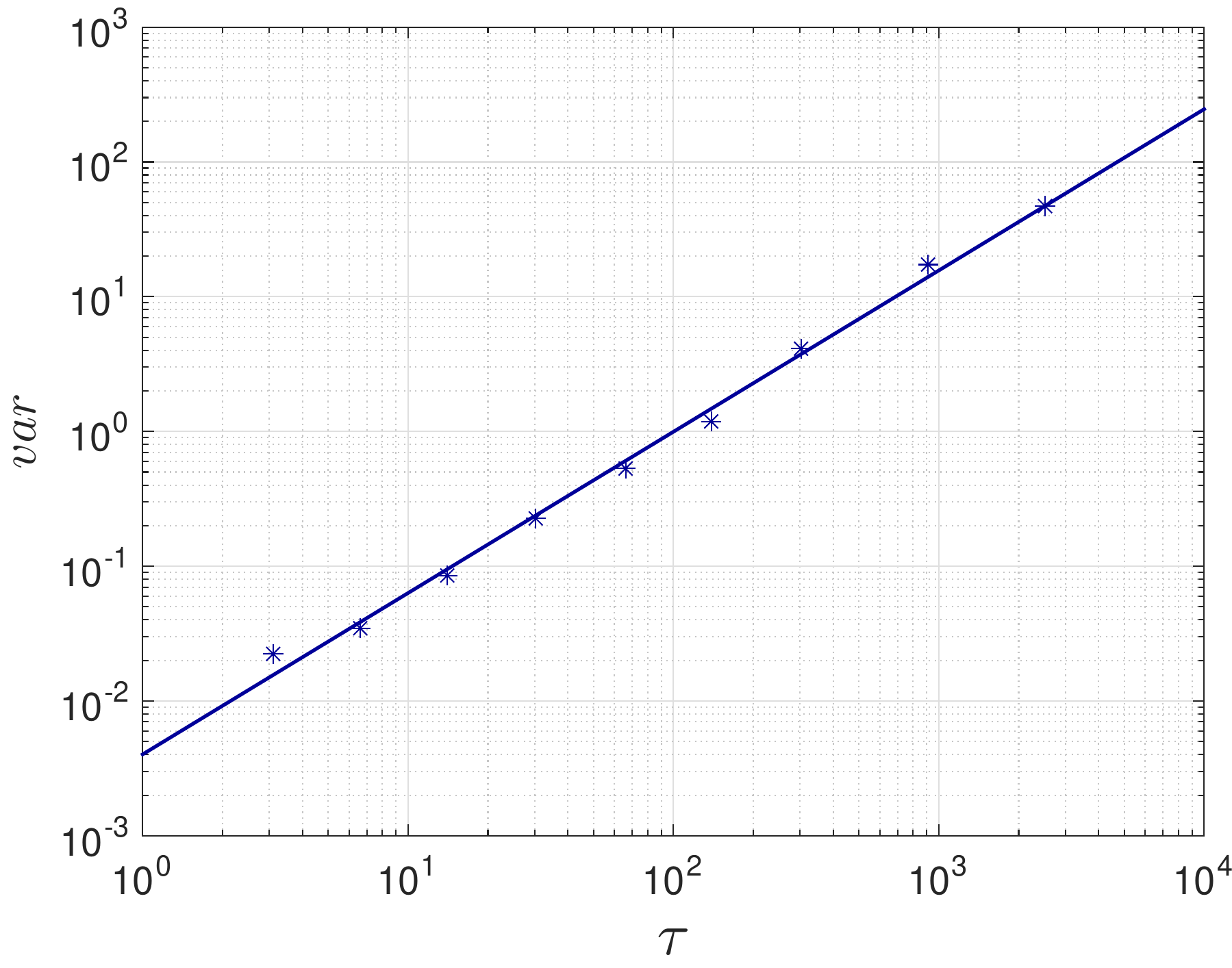}
  \caption{ Log-log plot of IMF variance as a function of  period for a fBm  of $H=0.6$ and length $T_1=10,000$. The blue line represents the least square fit. The scaling exponent  $H^*=0.593$ can be recovered from half the slope of the least square  fit.}
	  \label{fig:EMDH}
    \end{center}
	\end{figure}

Moreover, to visualize the linear relationship  of Equation \ref{eq:fBmScaling}, we explicitly show   the relation between $\log(var(IMF^{fBm}_k))$ and $\log(\tau_k)$ for a fBm simulation of  scaling exponent $H=0.6$ and length  $T_1=10,000$ points, see Figure ~\ref{fig:EMDH} . 
In this example, the  resulting estimator is $H^*=0.593$ which accurately approximates  the  scaling exponent of the simulated process.

\section{Variance scaling in intra-day financial data}
	
	\label{FinancialData}
	We analysed intra-day  prices for 22 different stocks market indices. The data set, extracted from Bloomberg, covers a period of 6 months from May  $5^{th}$, 2014 to  November  $5^{th}$, 2014. Prices are recorded at 30 second intervals\footnote{Prices for the  Warsaw stock exchange index (WIG) were only available at 
	 every minute frequency.}. 
	We excluded weekends and  holidays.  
	The number of working days and the number of points for every trading day depend on the opening hours of each stock exchange.  The list of analysed stock market indices is reported in Table \ref{tab:indexes}.

\begin{table}[h]
  \centering
  \resizebox{6.6 cm}{!} {
    \begin{tabular}{lcc}
   \hline
   	 	\hline
    \textbf{Country} & \textbf{Index} & \textbf{Length} \\
  \hline
  	 	\hline
    Brazil & BOVESPA &   105,000  \\
    China & SSE   &      60,480  \\
    France & CAC 40 &   136,080  \\
    Greece & ASE   &   106,470  \\
    Hong Kong & HSI   &      98,154  \\
    Hungary & BUX   &   122,880  \\
    Italy & FTSE MIB &   133,056  \\
    Japan & NIKKEI 225 &      75,600  \\
    Malaysia & KLSE  &   115,320  \\
    Mexico & IPC   &   100,620  \\
    Netherlands & AEX   &   130,680  \\
    Poland & WIG   &      64,680  \\
    Qatar & DSM   &      52,080  \\
    Russia & RTSI   &   133,120  \\
    Singapore & STI   &   123,840  \\
    South Africa & JSE   &   117,500  \\
    Spain & IBEX  &   135,527  \\
    Turkey & XU 100 &      91,760  \\
    UAE   & UAED  &      60,000  \\
    UK    & FTSE  &   130,560  \\
    USA   & S\&P 500 &      99,840  \\
    USA   & NASDAQ &   100,620  \\
  \hline
  	 	\hline 
    \end{tabular}%
    }
   	\caption{Stock market indices including the length of the time series.}
   		  \label{tab:indexes}%
\end{table}%

\begin{figure}
  \begin{center}
  \subfigure{
  \includegraphics[width=14.5 cm,  height=3 cm]{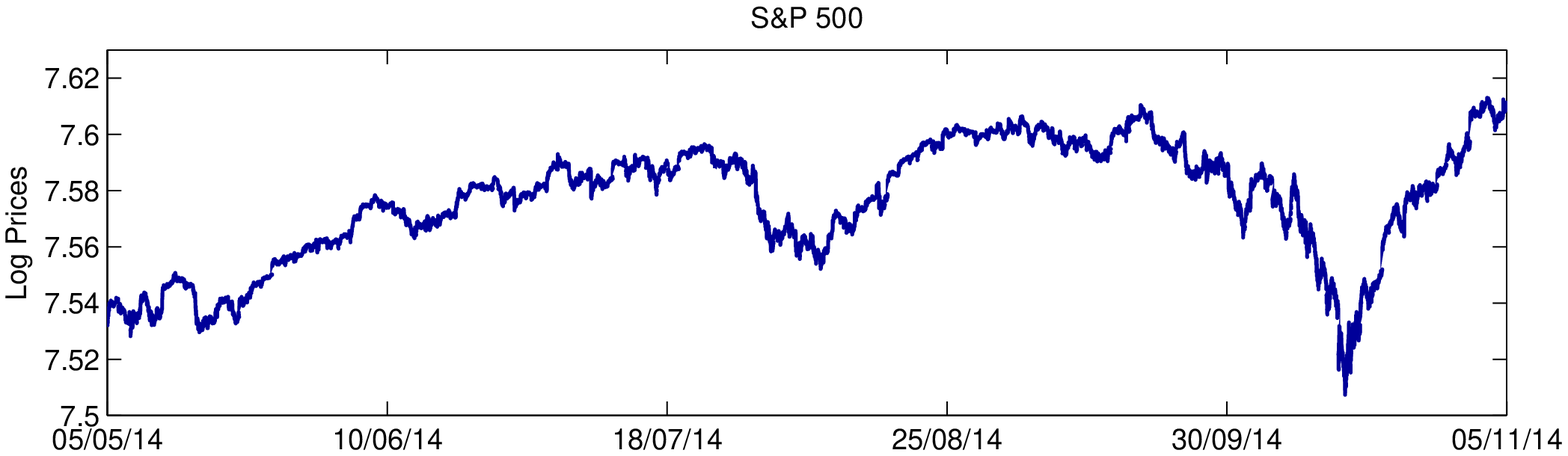}}
  \subfigure{
 \includegraphics[width=14.5 cm,  height=17 cm]{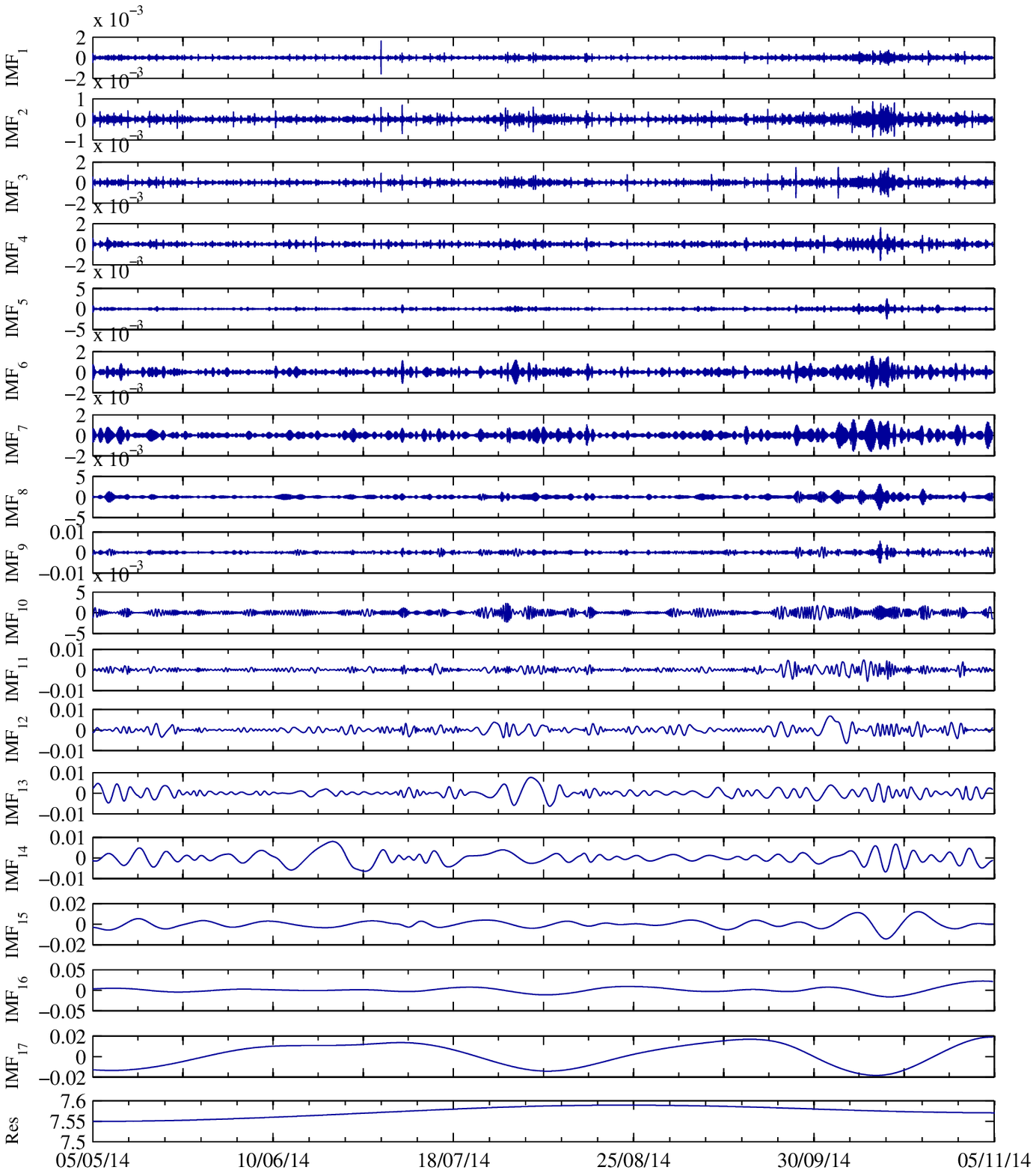}}
  \caption{Top: log-price time series of the S\&P 500 index for the period 05/05/2014 to 05/11/2014. Bottom: the  17 IMFs  and the residue obtained through EMD of the log-prices.
  \label{fig:SP}} 
  \end{center}
  \end{figure}

  We applied EMD to the logarithmic price of each financial time series. For the sake of clarity, in this section we only focus on the decomposition of the S\&P 500 index, but a similar analysis has been done for the other stock market indices. For the S\&P 500 log-price time series, we extracted 17 IMFs and a residue that describe the local cyclical variability of the original signal  and  represent it at different time-scales. The original log-price time series and its IMFs are displayed in Figure \ref{fig:SP}. 
  In this figure, we observe  temporary clusters of volatility that characterize some of the components,  for example the high volatility at the end of the time series can evidently be seen in components 2,3,4,6 and 7.

  The  IMF periods, calculated as the total number of data points divided by the total number of zero crossings, are reported in  Table \ref{tab:period}. These periods are converted into minutes, hours and days.
  The fastest component has a cycle of 1.6 minutes, contrasting the slowest cycle of  11.6 days. Notice that the first 12 IMFs represent the intra-day activity (6.5 hours of trading), while the remaining IMFs (from $13^{th}$ to $17^{th}$) are associated with the inter-day cycles. The last component is the residue of the EMD.

      \begin{table}[htbp]
                \centering
                \resizebox{11cm}{!} {
                 \begin{tabular}{cccccc}
              \hline
             	  			     \hline
             	       	        \textbf{IMF} & \textbf{Period/min} & \textbf{IMF} & \textbf{Period/hr} & \textbf{IMF} & \textbf{Period/days} \\
             	       	        \hline
             	       	        \hline
             	           \textbf{1} & 1.6   & \textbf{9} & 1.1   & \textbf{14} & 1.1 \\
             	           \textbf{2} & 2.8   & \textbf{10} & 1.9   & \textbf{15} & 2.2 \\
             	           \textbf{3} & 4.9   & \textbf{11} & 3.0   & \textbf{16} & 4.3 \\
             	           \textbf{4} & 8.4   & \textbf{12} & 5.9   & \textbf{17} & 11.6 \\
             	           \textbf{5} & 13.0  & \textbf{13} & 11.7  & \textbf{18} & Residue \\
             	           \textbf{6} & 19.3  &       &       &       &  \\
             	           \textbf{7} & 28.8  &       &       &       &  \\
             	           \textbf{8} & 41.7  &       &       &       &  \\    	       
             	       	        \hline
             	       	        \hline
             	       	        \end{tabular}%
             	       	        }          	       	    
             	       	       	      \caption{Period of the IMFs obtained from the  S\&P 500 index.}
             	       	       	  \label{tab:period}%
             	       	       	\end{table}%

\begin{figure}
  \begin{center}
             \includegraphics[width=12 cm,  height=3.5 cm]{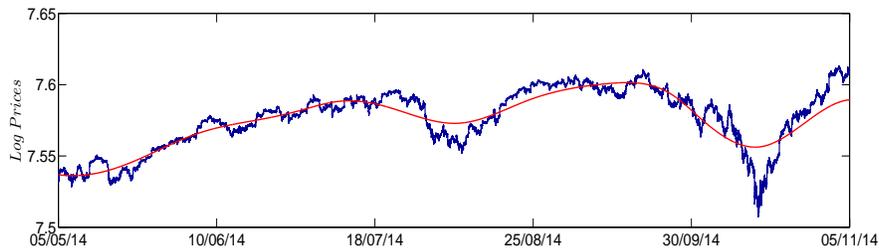} 
            \caption{Log-price time series of the S\&P 500 index (blue line). The red line represents the \lq trend\rq \, of the data calculated as the sum of the residue plus the  last IMF.} 
         \label{fig:trend}       
  \end{center}       	       	       
 	\end{figure}

The overall  trend of the time series is given by the residue, and each component can be seen as an oscillating trend of the previous component on a shorter time-scale. The effectiveness of EMD as a de-trending  and smoothing tool is illustrated in  Figure \ref{fig:trend}. In this figure, the original time series (blue line) is compared with a 'trend' (red line), calculated as the sum of the residue  plus the  last  component.

\begin{figure}
    \begin{center}
   	 \includegraphics[width=3.5 in]{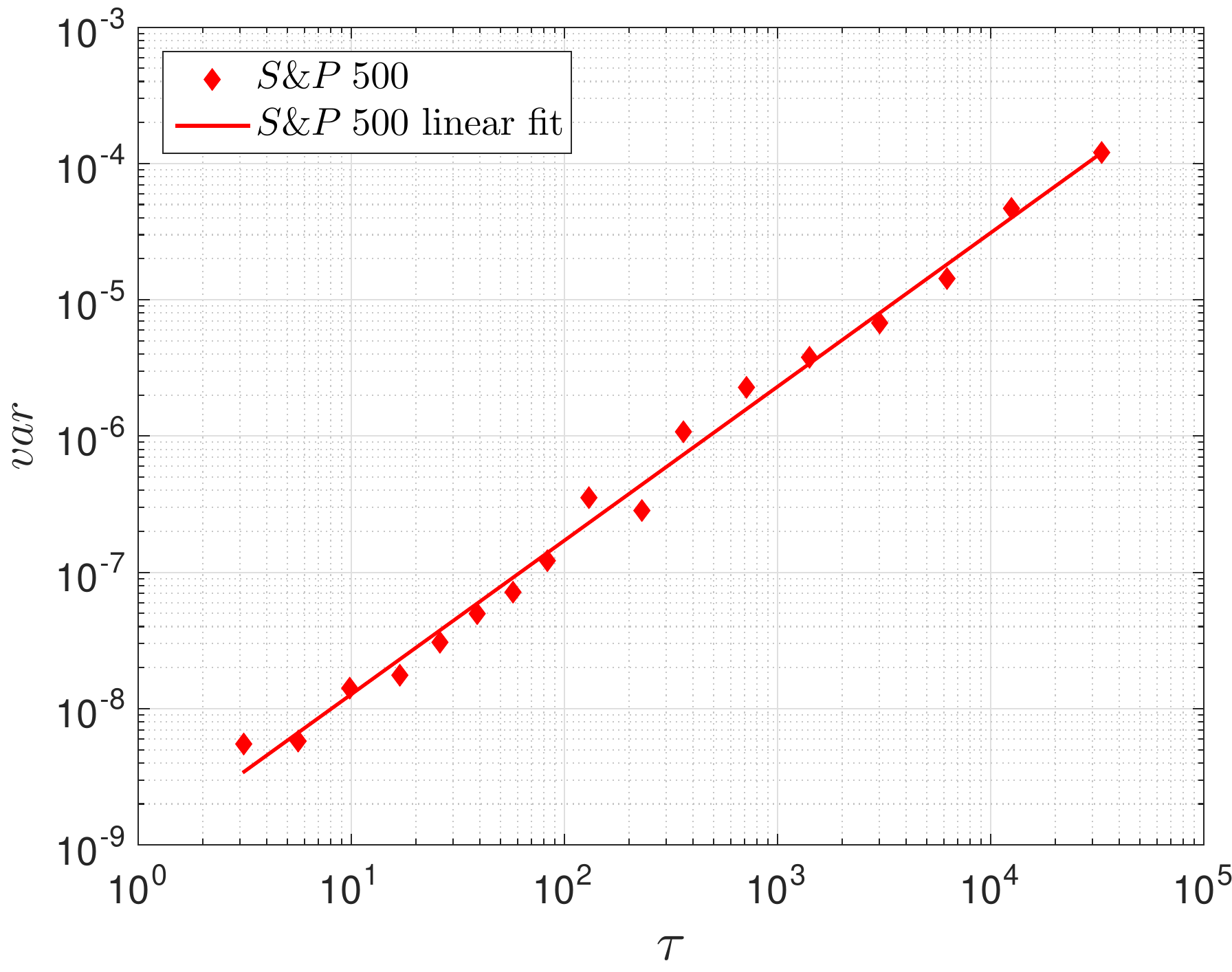} 
      \caption{Log-log plot of IMF variance as a function of  period for the EMD of the S\&P 500 index. The red line represents the best least square linear fit. The goodness of the linear fit is $R^2=0.992.$}     
   \label{fig:IMFAllSP}    
       \end{center}    	       	       
  	\end{figure}

 In previous section we discussed that for fBm, the EMD produces a linear relationship between the logarithmic values  of variance of the IMFs and its respective period of oscillation (Equation \ref{eq:fBmScaling}). We tested whether this relationship also holds for financial data.  In Figure \ref{fig:IMFAllSP}, we show the log-log plot of variance as a function of period for  the IMFs obtained from the  S\&P 500 index (red diamonds). The estimated  scaling exponent  has a value of $H^*=0.55$.  The goodness of the linear fit was estimated by the coefficient of determination\footnote{This coefficient of determination is the square of correlation between the dependent and independent variable. Values of  this coefficient range from 0 to 1, with 1 indicating a perfect fit between the data and the linear model, see for example \cite{rao73}.} which is $R^2=0.992$. 
We can conclude that this index satisfies the linear relationship of Equation \ref{eq:fBmScaling} but the scaling exponent, $H^*=0.55$, is different from that of Brownian motion, $ H=0.5$.

We performed the  same analysis for the other  stock  indices,  finding both significant deviations from  Brownian motion ($H^* \neq 0.5$ )  and  deviations from the scaling law of Equation \ref{eq:fBmScaling}.

In Table \ref{tab:Ranking1}, we report more details about the decomposition for each financial index. We include the number of   IMFs and the  index of orthogonality as described in Equation \ref{eq:IO}. We observe small values of the IO, indicating an almost orthogonal decomposition. Furthermore, we report the estimated exponent and the goodness of the linear fit for every stock market index.  
Although the coefficients of determination are all above 0.94,  we shall discuss shortly that significant deviations from linearity (fBm behaviour)  are observed, especially in less developed markets.

 \begin{table}[h!]
   \centering
     \resizebox{8.3cm}{!} {
     \begin{tabular}{ccccc}
     \hline
       \hline
     \textbf{ Index} & \textbf{\# IMFs}  &  \: \textbf{IO \footnotesize{$\times 10^{4}$}} \:  & \quad $ \boldsymbol{R^2} $ \quad &  \quad $\boldsymbol{H^*}$ \quad \\
     \hline
       \hline
    S\&P 500 & 17   & $3.7 $  & 0.992 & 0.564 \\
    BOVESPA & 18   & $6.3$  & 0.989 & 0.561 \\
    FTSE MIB & 18     & $8.6 $ & 0.987 & 0.571\\
    XU 100 & 19   & $3.5$  & 0.985 & 0.563 \\
    RTS   & 20   & $11$   & 0.985 & 0.581 \\
    CAC 40 & 17   & $8.8$  & 0.984 & 0.564 \\
    UAED  & 16     & $6.3$ & 0.978 & 0.616\\
    FTSE  & 23   & $2.9$  & 0.977 & 0.529 \\
    ASE   & 15    & $29$ & 0.974 & 0.587  \\
    IBEX  & 18     & $5.7$ & 0.973 & 0.531\\
    WIG   & 16    & $8.2$ & 0.973 & 0.591 \\
    SSE   & 14   & $1.6$  & 0.971 & 0.534 \\
    DSM   & 18   & $7.6$  & 0.971 & 0.618 \\
    IPC   & 18    & $0.68$ & 0.971 & 0.555 \\
    BUX   & 19     & $4.2$ & 0.970 & 0.542\\
    HSI   & 19    & $1.0$ & 0.969 & 0.554  \\
    AEX   & 21     & $13$ & 0.968 & 0.558 \\
    NASDAQ & 20   & $5.1$  & 0.960 & 0.530 \\
    NIKKEI 225 & 22   & $8.4$  & 0.959 & 0.544 \\
    JSE   & 19   & $2.9$  & 0.956 & 0.518 \\
    KLSE  & 22    & $0.77$  & 0.943 & 0.540 \\
    STI   & 21    & $2.6$ & 0.942 & 0.522 \\ 
     \hline
       \hline
     \end{tabular}%
     }
    \caption{\footnotesize Stock market indices including  the number of IMFs obtained when applying EMD to the logarithmic price. The second column report the index of orthogonality ($\times 10^{4}$). Stock market indices are reported in descending order of  $R^2$, which represents  the goodness of the linear fit of Equation \ref{eq:fBmScaling}. Last column reports the estimated  exponent $H^*$ of the same equation. 
    }
  \label{tab:Ranking1}%
 \end{table}%

Let us now discuss in more detail the deviations of the scaling laws found in stock markets  from the scaling expected in Brownian motion (Bm). 
With this aim, we generated  $N=100$  paths of Bm with length $T$ equal to the analysed stock market index (see Table \ref{tab:indexes}). We applied EMD to each  simulation and obtained its respective intrinsic oscillations denoted as  $IMF_k^{Bm_i}$, $i=1,2,\ldots,100$, $k=1,2,\ldots,n_i$, with $n_i$ the number of IMFs for each Bm simulation. In order to compare the variance of the IMFs extracted from the financial index $X$, $var(IMF_k^{X})$, against the  $var(IMF_k^{Bm_i})$,  we rescaled the latter as:

\begin{equation}
\widehat{var(IMF_k^{Bm_i})}= c_i \; var(IMF_k^{Bm_i}),
\end{equation}
where

\begin{equation}
c_i=\frac{\frac{1}{n}\sum\limits_{k=1}^{n}(var(IMF_k^{X})/\tau^{X}_k)}{\frac{1}{n_{i}}\sum\limits_{k=1}^{n_i}(var(IMF_k^{Bm_i})/\tau^{Bm_i}_k)}.
\end{equation}

\begin{figure}
 \begin{center}
    \includegraphics[width=3.5 in]{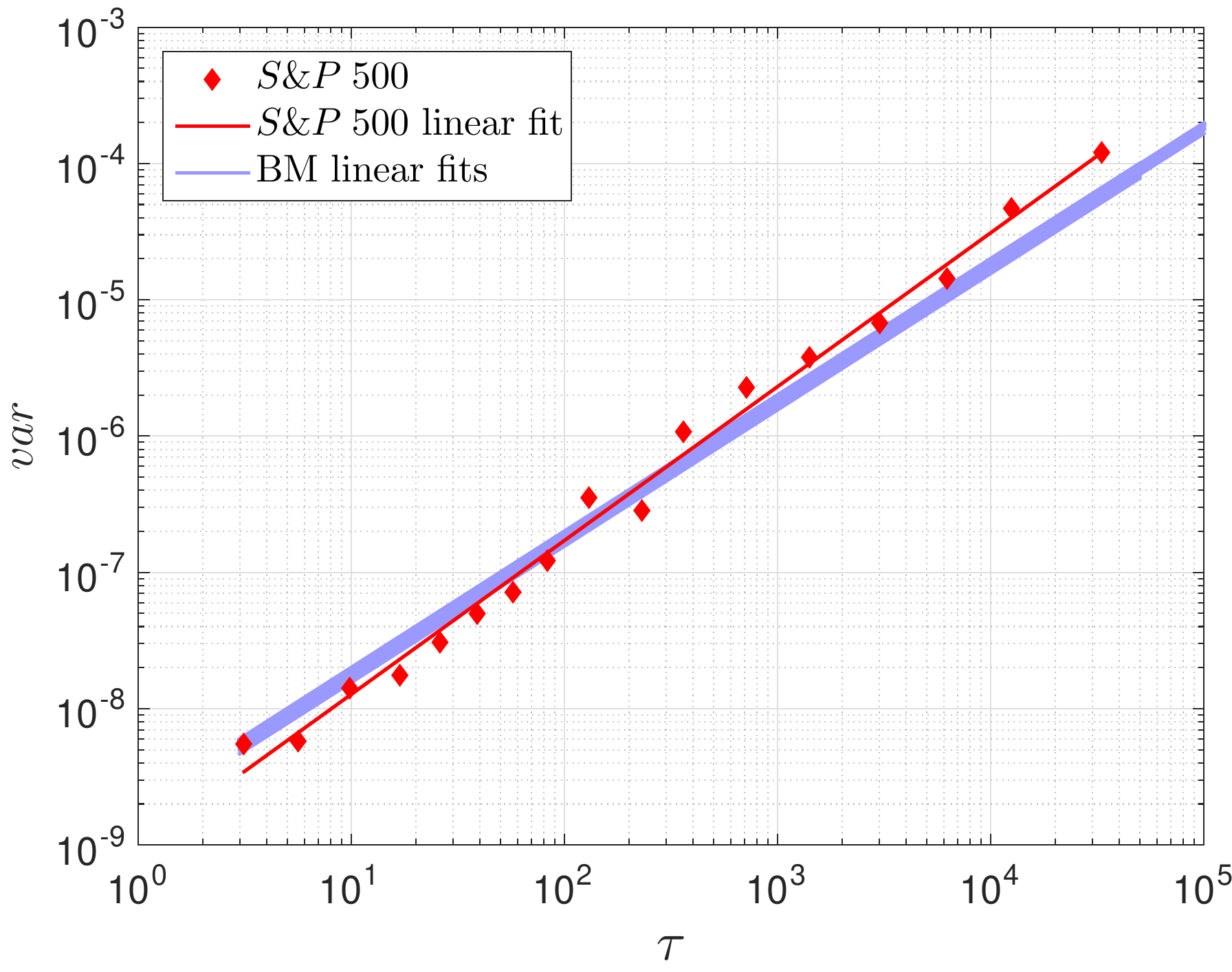}    
   \caption{Log-log plot of variance as a function of period for the  S\&P 500 IMFs (red diamonds) compared with 100 rescaled Bm linear fits of slope $H^*=0.5$ (blue lines).}  
    \label{fig:BMSP}  
     \end{center}               	          	         
    \end{figure}

For the rescaled $\widehat{var(IMF_k^{Bm_i})}$,  we estimated the intercept constant $c_{o_i}$  of Equation  \ref{eq:fBmScaling}, fixing $H=0.5$. In Figure  \ref{fig:BMSP}, we present all these 100 linear fits as light blue lines. In the same figure, we plotted the variance of the IMFs extracted from the S\&P 500 index, same  as reported in   Figure  \ref{fig:IMFAllSP}. 
We observe that the Brownian motion linear fits (blue lines) and the linear fit of the  S\&P 500 index (red line) are  close to each other, suggesting an efficient behaviour in this market. 

The  goodness of the linear fit between the financial data points (red diamonds) and   each of the  Brownian motion linear fit (blue lines) was calculated as follow:

\begin{equation}
R^2_{Bm_i}= 1-\frac{\sum\limits_{k=1}^{n}\left[\log\left(var\left(IMF_k^X\right)\right)- \log\left(c_i\;c_{0_i} \tau_k^X \right)\right] ^2}{\sum\limits_{k=1}^{n}\left[\log\left(var\left(IMF_k^X\right)\right)- \left<\log\left(var\left(IMF_k^X\right)\right)\right> \right]^2}.
\label{eq:R2}
\end{equation}
 The  number of IMFs extracted from the stock index $X$ is denoted by $n$ and $\left<\cdot\right>$ indicates the mean over these $n$ IMF variances. 
 The deviations from Bm were calculated by the mean over the goodness of the linear fits, i.e. we calculated $\left<R_{Bm}^2\right>=\frac{1}{n}\sum\limits_{i=1}^{100}R^2_{Bm_i}$.
 For the S\&P 500 index, we obtained a coefficient equal to  $\left<R_{Bm}^2\right>=0.979$, demonstrating the similarity between the scaling properties of this financial index and Brownian motion.

\subsection{ Scaling properties of stock market indices}
In previous section, we introduced  two measures that quantify the  deviations from the scaling behaviour of fractional Brownian motion and   Brownian motion. These measures are given by:
\begin{enumerate}
\item $R^2$, coefficient of determination (square of correlation) between the logarithm of period and logarithm of variance of IMFs obtained from stock market indices.
\item $\left<R_{Bm}^2\right>$,  mean of the relative squared residuals between the  IMF variances obtained from financial data and each of the linear fits for  Brownian motion simulations.

\end{enumerate} 
In Table \ref{tab:Ranking}, we report the values of $\left<R_{Bm}^2\right>$  for the 22 stock  market indices. For comparison purposes, we repeated the $R^2$ values. The last column in this table indicates the ordering of the markets if $R^2$ were used as the ranking measure.

\begin{table}[htbp]
  \begin{center}
  \resizebox{12cm}{!} {
    \begin{tabular}{cccccc}
   \hline
   \hline
   \textbf{Country} & \textbf{Index} &  $\boldsymbol{\left<R_{Bm}^2\right>}$ & $\boldsymbol{Rank_{\left<R_{Bm}^2\right>}}$ & $\boldsymbol{R^2}$ & $\boldsymbol{Rank_{R^2}}$ \\
    \hline
    \hline
       USA   & S\&P 500 & 0.979 & \textbf{1} & 0.992 & \textbf{1} \\
       Brazil & BOVESPA & 0.977 & \textbf{2} & 0.989 & \textbf{2} \\
       UK    & FTSE  & 0.973 & \textbf{3} & 0.977 & \textbf{8} \\
       Turkey & XU 100 & 0.972 & \textbf{4} & 0.985 & \textbf{4} \\
       Italy & FTSE MIB & 0.971 & \textbf{5} & 0.987 & \textbf{3} \\
       France & CAC 40 & 0.970 & \textbf{6} & 0.984 & \textbf{6} \\
       Spain & IBEX  & 0.969 & \textbf{7} & 0.973 & \textbf{10} \\
       China & SSE   & 0.967 & \textbf{8} & 0.971 & \textbf{12} \\
       Russia & RTSI   & 0.964 & \textbf{9} & 0.985 & \textbf{5} \\
       Hungary & BUX   & 0.963 & \textbf{10} & 0.970 & \textbf{15} \\
       Mexico & IPC   & 0.960 & \textbf{11} & 0.971 & \textbf{14} \\
       Hong Kong & HSI   & 0.958 & \textbf{12} & 0.969 & \textbf{16} \\
       USA   & NASDAQ & 0.954 & \textbf{13} & 0.960 & \textbf{18} \\
       Netherlands & AEX   & 0.953 & \textbf{14} & 0.968 & \textbf{17} \\
       South Africa & JSE   & 0.952 & \textbf{15} & 0.956 & \textbf{20} \\
       Japan & NIKKEI 225 & 0.949 & \textbf{16} & 0.959 & \textbf{19} \\
       Greece & ASE   & 0.948 & \textbf{17} & 0.974 & \textbf{9} \\
       Poland & WIG   & 0.947 & \textbf{18} & 0.973 & \textbf{11} \\
       UAE   & UAED  & 0.939 & \textbf{19} & 0.978 & \textbf{7} \\
       Singapore & STI   & 0.934 & \textbf{20} & 0.942 & \textbf{22} \\
       Malaysia & KLSE  & 0.933 & \textbf{21} & 0.943 & \textbf{21} \\
       Qatar & DSM   & 0.928 & \textbf{22} & 0.971 & \textbf{13} \\  
    \hline
    \hline
    \end{tabular}%
    }
    \caption{Stock market indices ranked in descending order of $\left<R_{Bm}^2\right>$. The last column indicates the ordering of the markets with respect to $R^2$.}
  \label{tab:Ranking}%
   \end{center}
\end{table}%

\begin{figure}
   \begin{center}
          \includegraphics[width=14 cm]{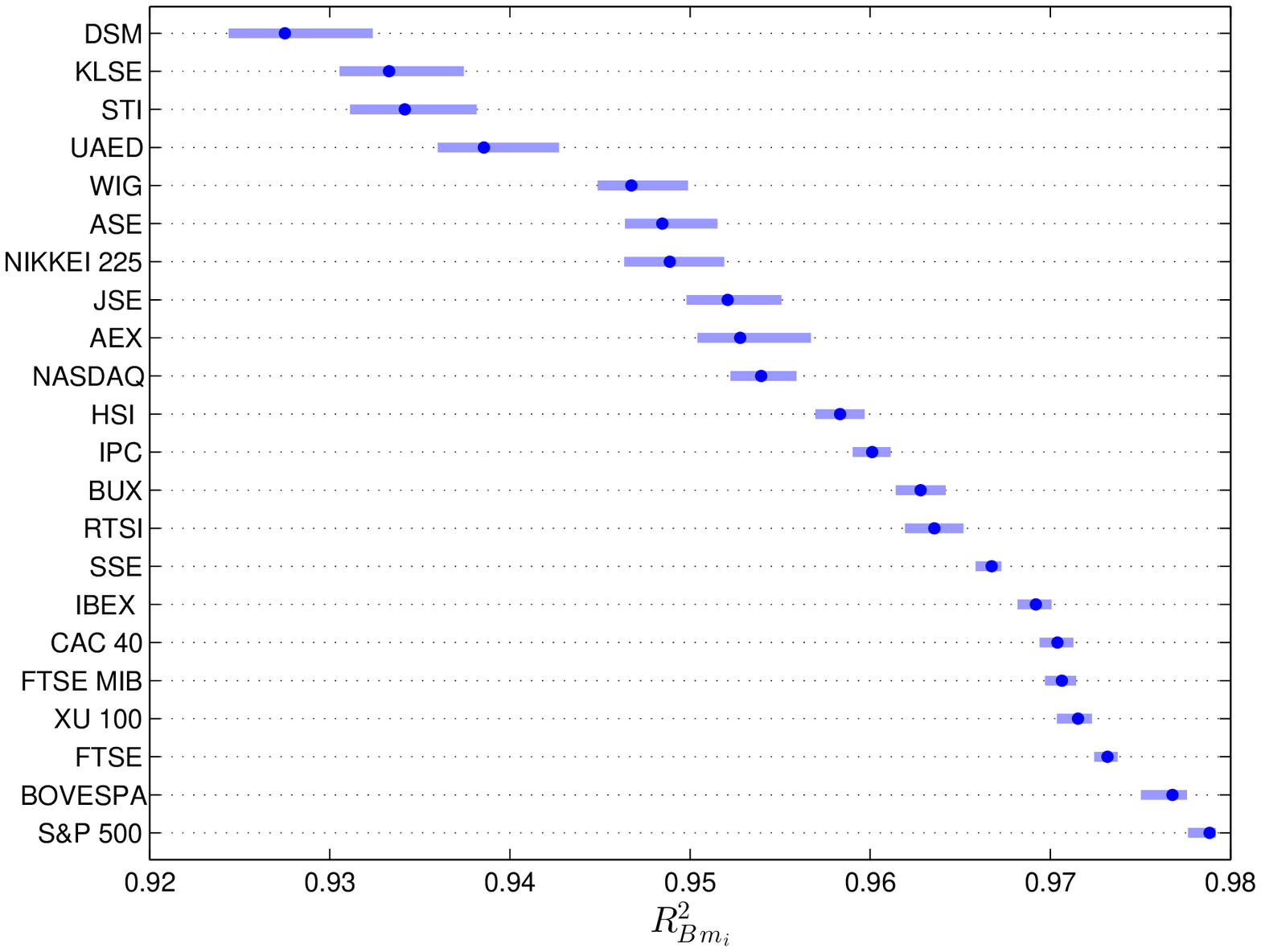}                       
     	 \caption{ Percentiles   $5^{th}$ and $95^{th}$ of the $R_{Bm_i}^2$ distribution  for the analysed stock market indices. The blue dot inside each bar indicates the value of  $\left<R_{Bm}^2\right>$ used for the financial market ranking.}  
           \label{fig:INDICES}    
           \end{center}   	   
         \end{figure}

The S\&P 500 index is ranked the highest in both scales. Developed markets tend to be at the  top of the  table with some exceptions that may arise from the specific characteristics of the  analysed period of time,  May  $5^{th}$, 2014 to  November  $5^{th}$, 2014. 
In Figure \ref{fig:INDICES}, we plot the financial market ranking. The horizontal bars represent  the $5^{th}$ and $95^{th}$ percentiles of the $R^2_{Bm_i}$ distribution. The blue dot inside each bar indicates the mean value $\left<R_{Bm}^2\right>$ as reported in Table \ref{tab:Ranking}. 
Despite the fact that some  financial stock indices have similar values of  $R^2_{Bm_i}$,  we can recognize statistically significant differences between developed and emerging markets, observing a clear tendency for the developed markets to present larger values of $\left<R_{Bm}^2\right>$ with  narrower distributions.

In order  to visualize the anomalous scaling in some stock markets and to understand the origin of the differences in the results,  we compare  the cases of the   NASDAQ (USA), BOVESPA (Brazil),	 NIKKEI 225 (Japan) and  DSM (Qatar) indices in more detail.

\begin{figure}
  \begin{center}
  \begin{minipage}{135mm}
  \subfigure[]{
  \resizebox*{6.5cm}{!}{\includegraphics{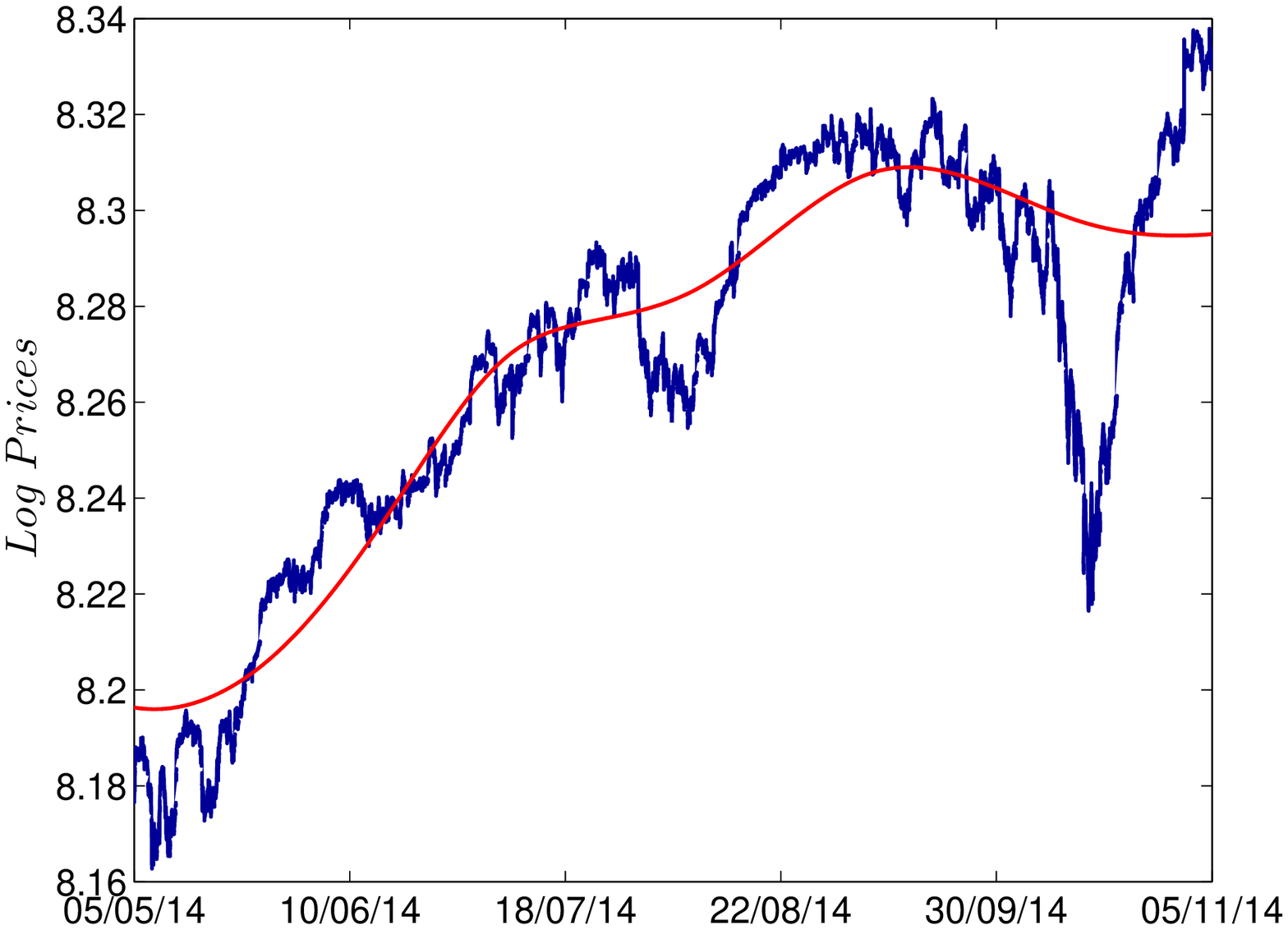}}\label{fig:NASDAQA}}
  \subfigure[]{
  \resizebox*{6.5cm}{!}{\includegraphics{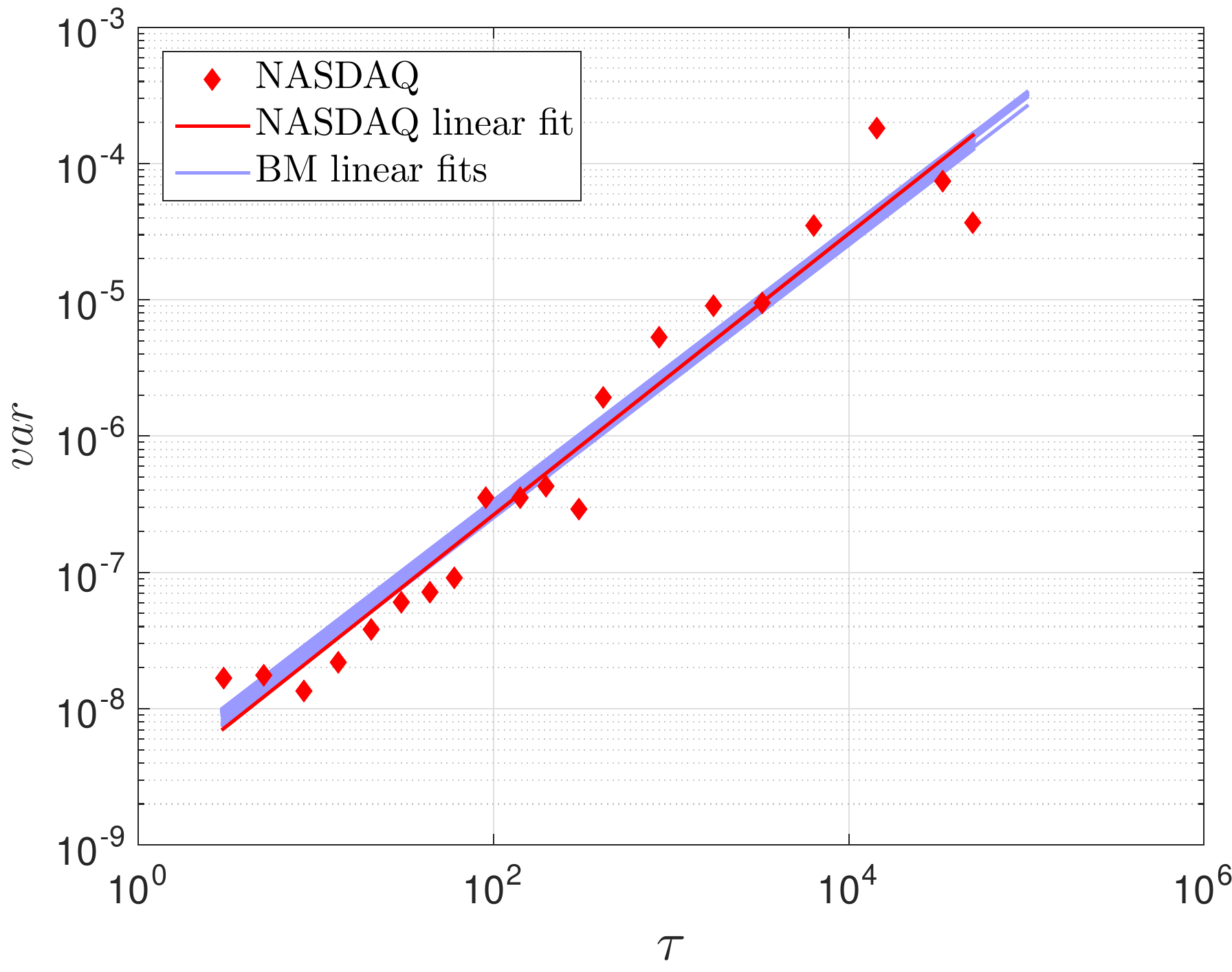}}\label{fig:NASDAQB}}
  \caption{EMD analysis for the NASDAQ index. Captions for figures (a) and (b)  are the same as captions for figures  (4) and (6) respectively.
  \label{fig:NASDAQ}}
  \end{minipage}
  \end{center}
  \end{figure}

For the NASDAQ index (USA), we obtained 20 IMFs and a residue. In Figure \ref{fig:NASDAQA}, we present the  log-price time series (blue line) and the \lq trend\rq \, consisting of the residue plus the last IMF (red line).   In  Figure \ref{fig:NASDAQB}, we observe that the deviation from the linear relationship of Equation \ref{eq:fBmScaling} is significant. Thus,  the log-log relationship between period and variance  is not completely satisfied. The resultant coefficient of determination is $R^2= 0.960$, ranking this index at the $18^{th}$ position.
 Moreover, when compared with Bm, we  identify that most of the components deviate  from the Bm  linear fits (blue lines).  We also note that the total number of component (21) is considerable larger than what would be expected from a process with uniform scales, i.e., $\log_2(100620)$=16.6. The presence of these extra oscillations with reduced variance suggests a more complex structure than Bm. The deviations from Bm model, quantified by the coefficient   $\left<R_{Bm}^2\right>=0.958$, rank this index at the $13^{th}$ position.

\begin{figure}
  \begin{center}
  \begin{minipage}{135mm}
  \subfigure[]{
  \resizebox*{6.5cm}{!}{\includegraphics{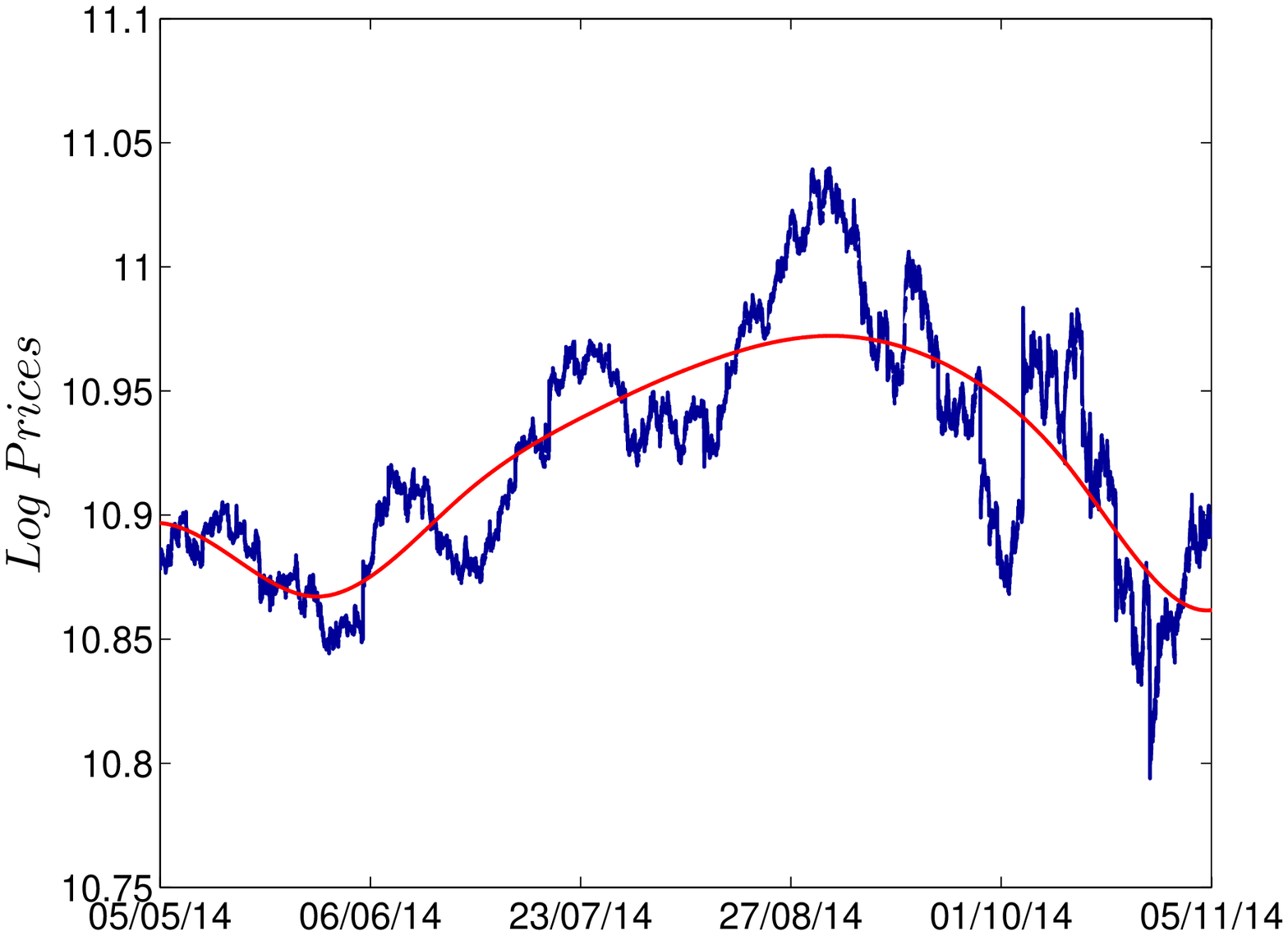}}\label{fig:BOVESPAA}}
  \subfigure[]{
  \resizebox*{6.5cm}{!}{\includegraphics{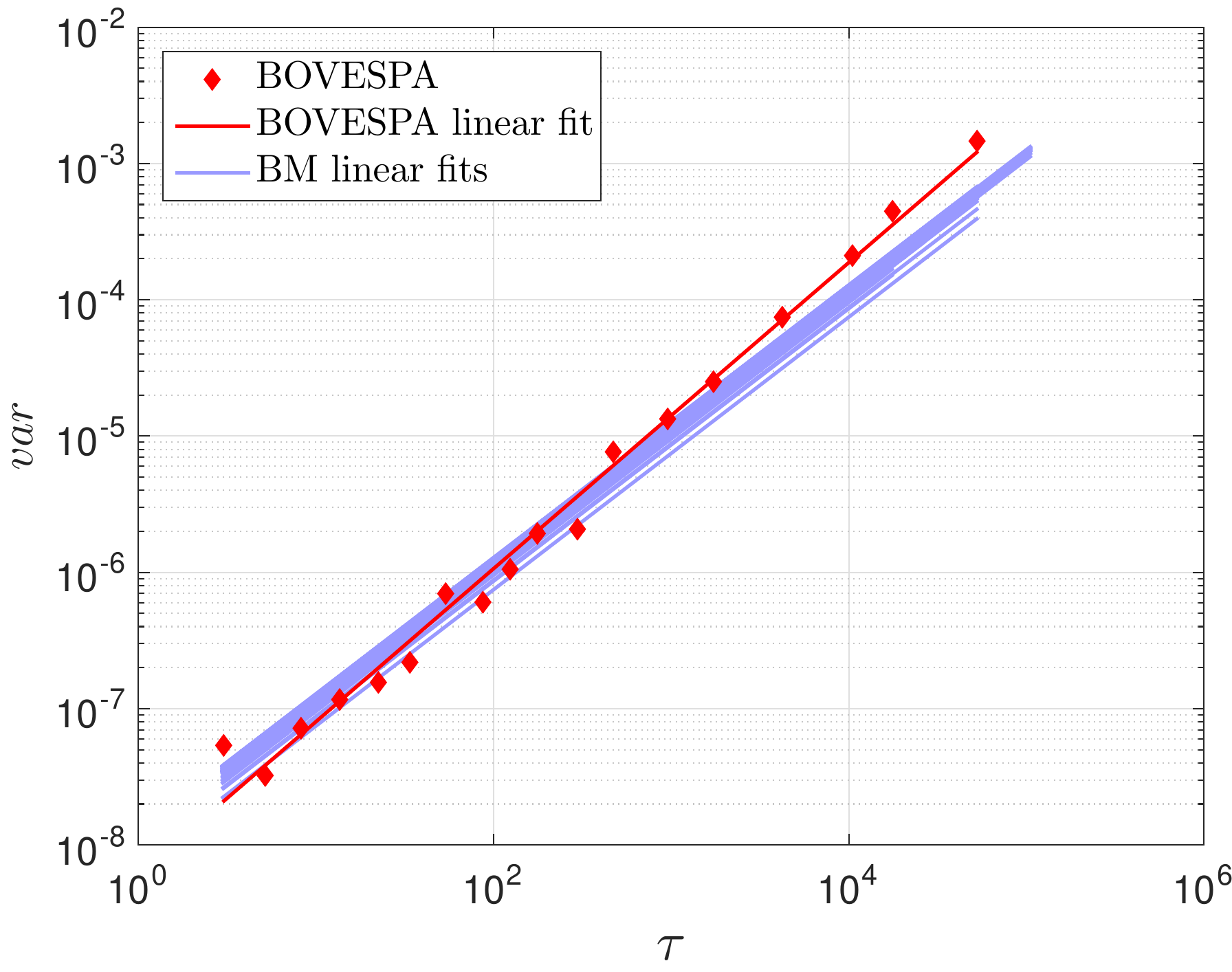}}\label{fig:BOVESPAB}}
  \caption{EMD analysis for the BOVESPA index. Captions for figures (a) and (b)  are the same as captions for figures  (4) and (6) respectively.
  \label{fig:BOVESPA}}
  \end{minipage}
  \end{center}
  \end{figure}

The variance scaling properties of the BOVESPA index (Brazil) are presented in  Figure \ref{fig:BOVESPA}.  For this stock index, the EMD identifies long period cycles with larger variance than what would be expected from Bm, see Figure \ref{fig:BOVESPAB}. However, the  linear fit between the logarithmic value of IMF variances and periods is in general good with $R^2=0.989$. The goodness of the linear fit between the Bm simulations is $\left<R_{Bm}^2\right>=0.977$, placing this index at the second position. Such a good ranking for this market may be unexpected, but we must stress that it only reflects the six-month period of observations. From Figure \ref{fig:BOVESPAA}, we can see that this was a rather random but calm period.

\begin{figure}
  \begin{center}
  \begin{minipage}{135mm}
  \subfigure[]{
  \resizebox*{6.5cm}{!}{\includegraphics{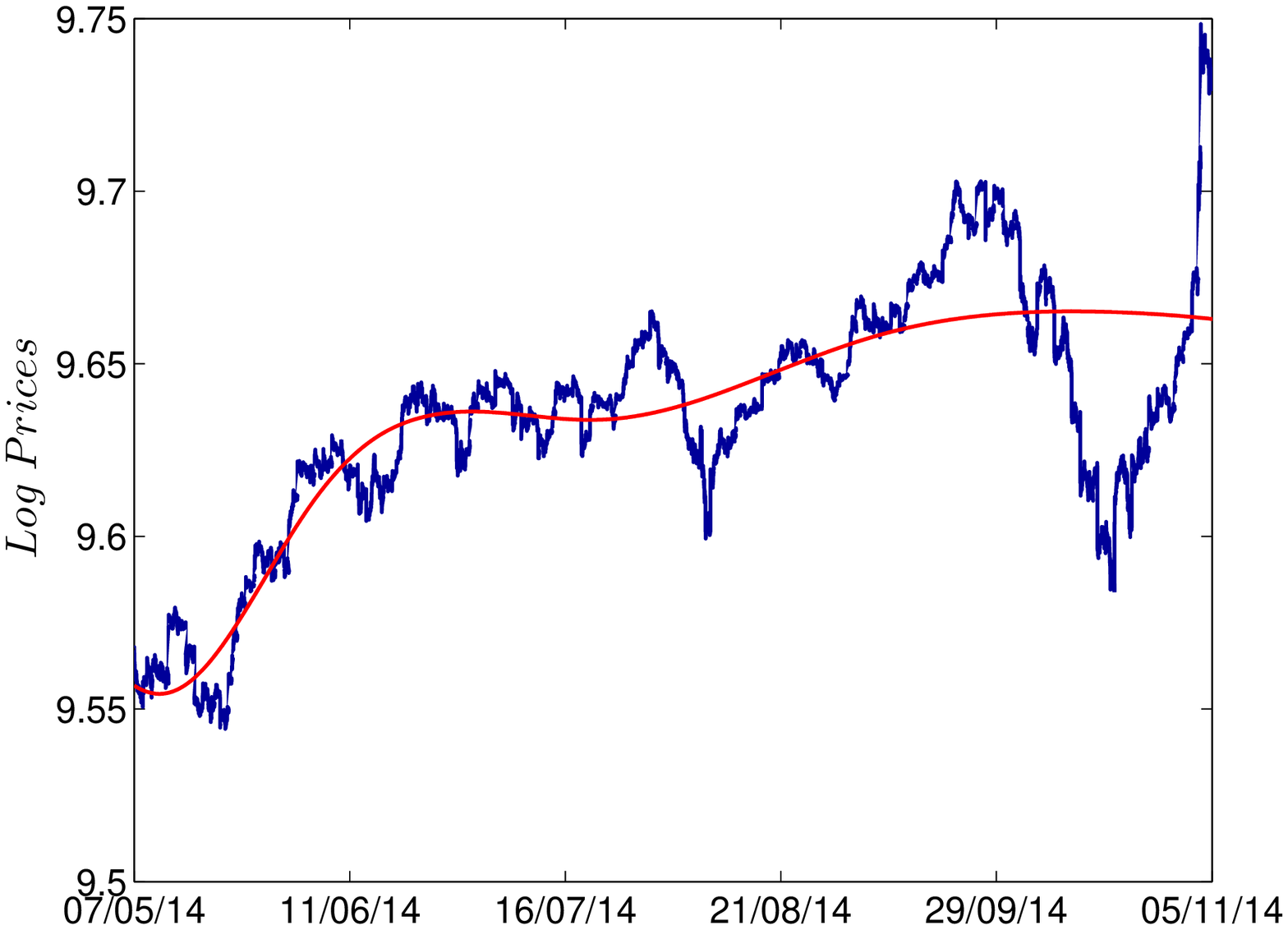}}\label{fig:NIKKEIA}}
  \subfigure[]{
  \resizebox*{6.5cm}{!}{\includegraphics{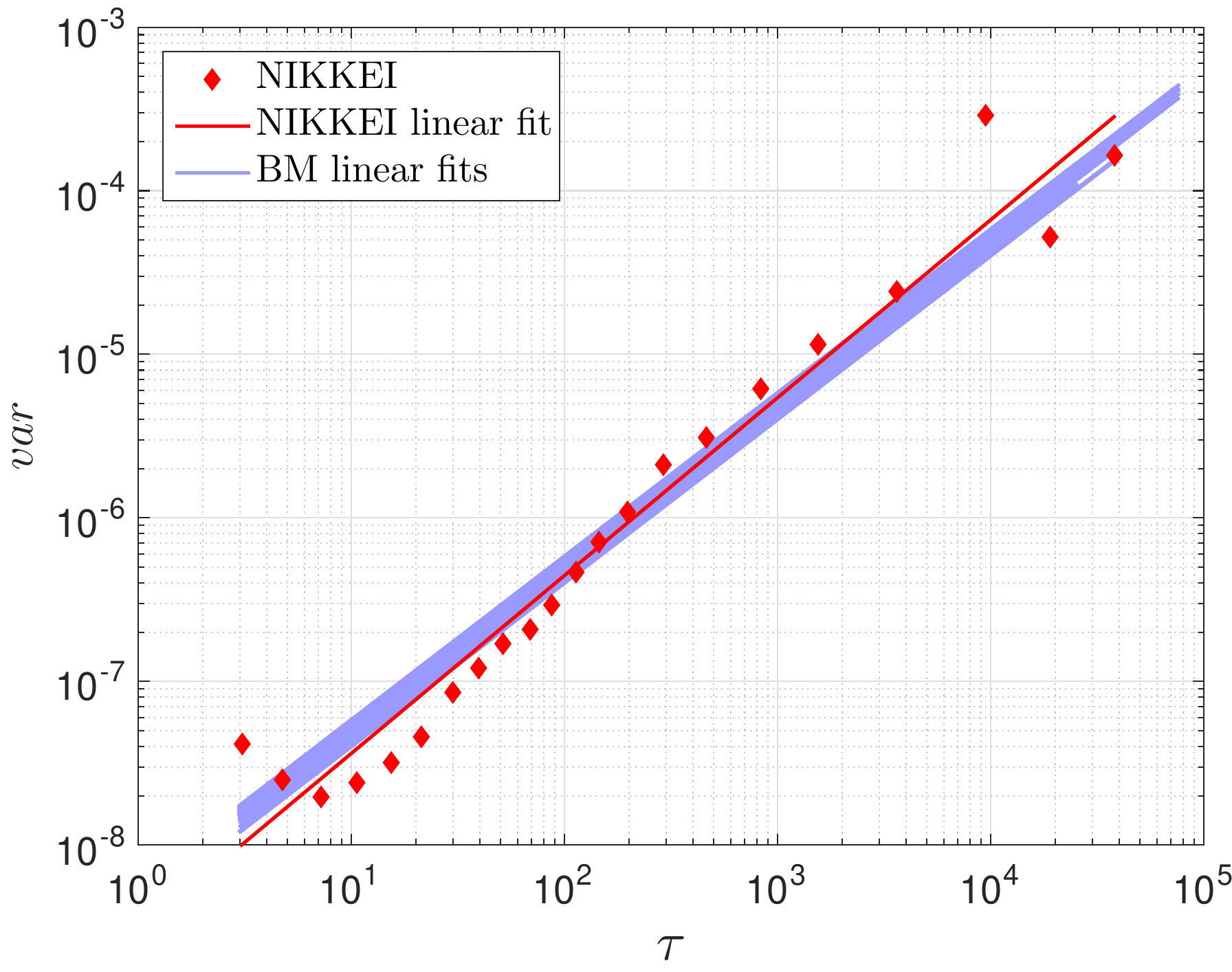}}\label{fig:NIKKEIB}}
  \caption{EMD analysis for the NIKKEI 225 index.  Captions for figures (a) and (b)  are the same as captions for figures  (4) and (6) respectively.
  \label{fig:NIKKEI}}
  \end{minipage}
  \end{center}
  \end{figure}

For the NIKKEI 225 index (Japan), we obtained 22 IMFs and a residue. Similar as the NASDAQ index, the number of components  is considerable larger than what would be expected from Bm, i.e., $\log_2(75600)$=16.2.
These many oscillations, specially the high frequency components, generate a non-linear behaviour that deviates from  Bm. Given the anomalous scaling behaviour of this stock index, see Figure \ref{fig:NIKKEIB}, we obtained $\left<R_{Bm}^2\right>=0.949$, ranking it at the $16^{th}$ position. 
 
\begin{figure}
   \begin{center}
   \begin{minipage}{135mm}
   \subfigure[]{
   \resizebox*{6.5cm}{!}{\includegraphics{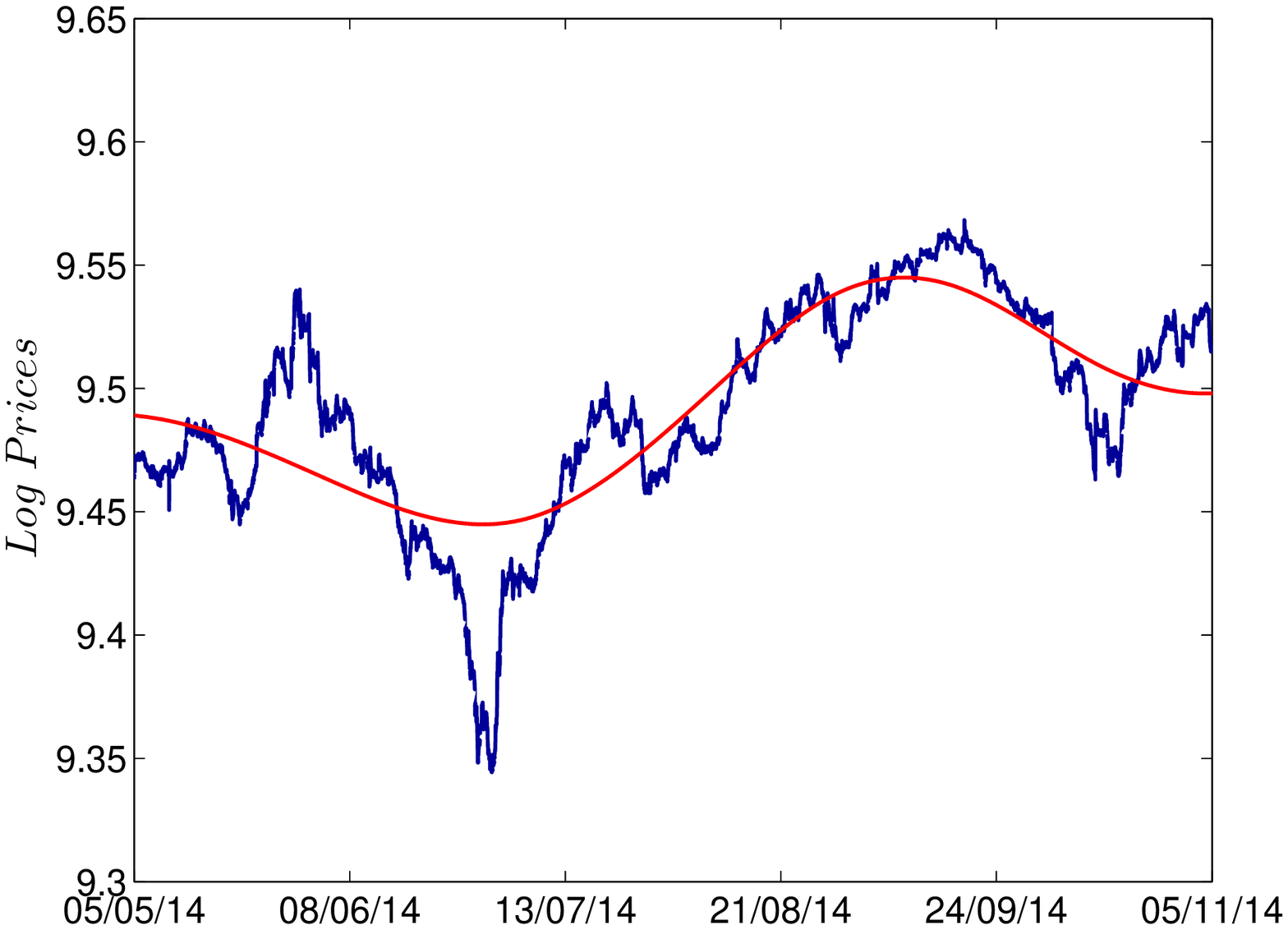}}\label{fig:QATARA}}
   \subfigure[]{
   \resizebox*{6.5cm}{!}{\includegraphics{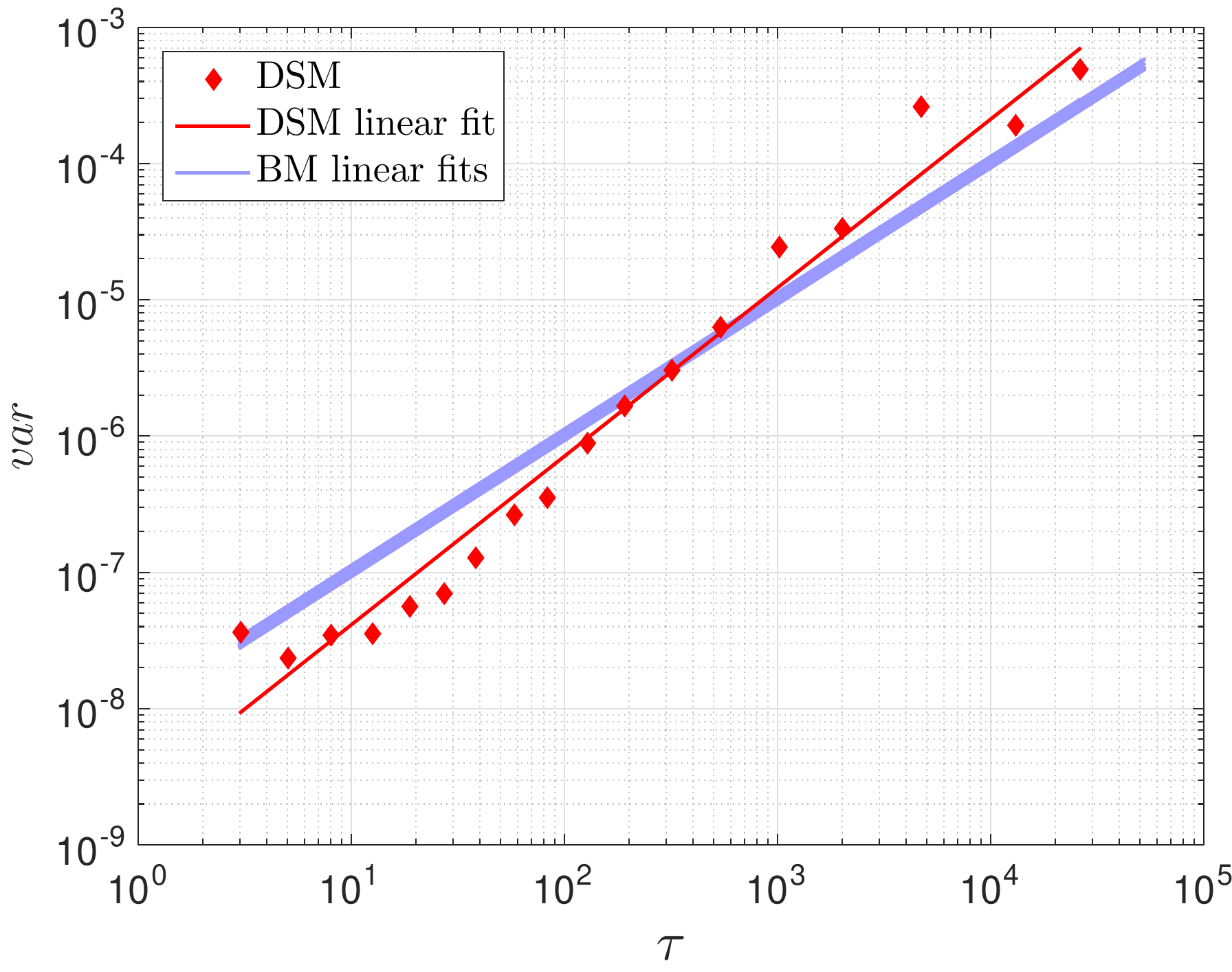}}\label{fig:QATARB}}
   \caption{EMD analysis for the DSM index.  Captions for figures (a) and (b)  are the same as captions for figures  (4) and (6) respectively.
   \label{fig:QATAR}}
   \end{minipage}
   \end{center}
   \end{figure}

 Finally,  the DSM stock index (Qatar) is displayed in Figure \ref{fig:QATAR}. The log-price time series and its respective \lq trend\rq  \,are displayed in Figure \ref{fig:QATARA}. 
 In Figure \ref{fig:QATARB}, we observe the poor liner fit of Equation \ref{eq:fBmScaling} that is characterized by a considerable steep slope. We obtained $R^2=0.971$, ranking this index at the lowest position.  Furthermore, if we compare  its  IMF variances against the Bm linear fits,  we observe that most of the variance values (red diamonds) follow outside the band expected from Bm. 
 The large variance of the  low frequency components suggests the presence of important long period cycles. Given its deviations from Bm behaviour, this index is also ranked the lowest with respect to the measure $\left<R_{Bm}^2\right>=0.928$.
 
\section{Conclusions}
\label{conclusions}
We explored the scaling properties of EMD, an algorithm that detrends and  separates  time series into a set of oscillating components called IMFs which are associated with specific time-scales.
We empirically showed  that  fBm obeys a scaling law that relates linearly the logarithm of the variance and the logarithm of the period of the IMFs. For fBm, we demonstrated that the extracted  coefficient of proportionality equals  the scaling exponent $H$ multiplied by two. 
When applied to  stock market indices,  the EMD reveals instead different scaling laws that can deviate significantly from both Brownian motion and fractional Brownian motion behaviour.
In particular, we noted that the EMD of high frequency financial data results in  a larger number of IMFs than what would be expected from Brownian motion. These many components, specially with high frequencies, create a curvature that disobeys the   linearity in the log-log relation between IMF variance and period found in fBm. This is a direct indication of anomalous scaling that reveals a  more complex structure in financial data than in self-similar processes. 

In this study, we applied  EMD to 22 different stock indices and  observed  that developed markets (European and North American markets) tend to have scaling properties closer to Brownian motion properties.  Conversely, larger deviations from uni-scaling laws are observed in some emerging markets  such as Malaysian and Qatari. 
	
These findings are in agreement with the discernible characteristics of developed and emerging markets, the former type being more likely to exhibit an  efficient behaviour, see for example \cite{DiMatteo,Matteo2005}.
Compared to previous approaches, the EMD method has the advantage to directly quantify  the cyclical components with strong deviations, giving a further instrument to understand the origin of market inefficiencies.
 	
\section*{Acknowledgements}
The authors wish to thank Bloomberg for providing the data. NN would like to acknowledge the financial support from CONACYT-Mexico. TDM wishes to thank the COST Action TD1210 for partially supporting this work. TA acknowledges support of the UK Economic and Social Research Council (ESRC) in funding the Systemic Risk Centre [ES/K002309/1].

	\bibliographystyle{elsarticle-num}

	\end{document}